# Emerging Technologies and Research Challenges for 5G Wireless Networks

Woon Hau Chin, Zhong Fan, and Russell Haines
Toshiba Research Europe Limited, Bristol, BS1 4ND, United Kingdom

Email: {woonhau.chin, zhong.fan, russell.haines}@toshiba-trel.com

## Abstract

As the take-up of Long Term Evolution (LTE)/4G cellular accelerates, there is increasing interest in technologies that will define the next generation (5G) telecommunication standard. This paper identifies several emerging technologies which will change and define the future generations of telecommunication standards. Some of these technologies are already making their way into standards such as 3GPP LTE, while others are still in development. Additionally, we will look at some of the research problems that these new technologies pose.

## Introduction

The mobile phone has evolved rapidly over the past decade from a monochrome device with a minuscule screen and little processing power to one with high resolution, palm sized screen and processing power rivalling a laptop. This transformation, coupled with an expanding cache of bandwidth hungry applications have triggered demands for higher data rates. Mobile data traffic has been forecasted to grow more than 24-fold between 2010 and 2015, and more than 500-fold between 2010 and 2020 [1]. This has in turn propelled the uptake of 4G contracts and driven operators worldwide to deploy 4G networks.

As the dust around 4G begins to settle, attention is now slowly turning towards future 5G technologies. A key feature of 4G, is its ability to support high data rate (up to 1 Gbit/s) on the downlink. However, while 5G will no doubt continue to up the ante on the data rate, we believe that the focus will also be on coverage and user experience. There are already simmering interests in beyond 4G technologies, and the industry is starting to fund projects looking into such technologies. However, the industry's view of the future wireless standard is mostly focused on data rates and efficiency, with heavyweights such as Qualcomm and Nokia Siemens Networks looking at technologies which will cope with traffic growth of 1000 times.

While there is no industry consensus on what 5G will ultimately be, apart from the usual higher data rate and energy efficiency, there are some emerging signs of things to come. For example, in the new IEEE 802.11 High Efficiency Wireless (HEW) study group, there is a pronounced increase in the presence of cellular operators, something not previously seen. This indicates growing interests to amalgamate different technologies to support future connectivity and data rates. Therefore, we believe that 5G will consist of multiple interconnected communication standards, ranging from wireless metropolitan area networks down to wireless personal networks, providing the required throughput and connectivity.

In this article, we identify several technologies, ranked in perceived importance, which will be crucial in future wireless standards. These may improve capacity, coverage, or energy efficiency. We have classified some of these technologies according to what they are trying to achieve in Table 1. Table 2 shows another taxonomy of these technologies in terms of their functionalities in the network. We also present the status of some of the wireless standardization bodies, including 3GPP, IEEE 802.11, and OneM2M and identify some of the research challenges that these technologies will bring.

## 1. Heterogeneous Networks

### Small Cells

As the demand for higher data rates increases, one of the solutions available to operators is to reduce the size of the cell. By reducing the size of the cell, area spectral efficiency is increased through higher frequency reuse, while transmit power can be reduced such that the power lost through propagation will be lower. Additionally, coverage can be improved by deploying small cells indoors where reception may not be good and offloading traffic from macro cells when required. This solution has only been made possible in recent years with the advancement in hardware miniaturization and the corresponding reduction in cost. Additionally, changes to the functional architecture of the access network allowed data and control signals to tunnel through the Internet, enabling small cells to be deployed anywhere with Internet connectivity. Small cells can have different flavors, with low powered femtocells typically used in residential and enterprise deployments, and the higher powered picocells used for wider outdoor coverage or filling in macro cell coverage holes.

The concurrent operation of different classes of base stations, macro-, pico-, and femto- base stations, is known as heterogeneous networks (or HetNets). This is used to provide a flexible coverage area and improve spectral efficiency. Overlaying different classes of base stations can also potentially provide a solution for the growing data traffic, especially when the transport of data is optimized to take advantage of the characteristics of heterogeneous networks. 3GPP has identified various scenarios and requirements for the enhancement of small cells in [2].

### New Carrier Type

One of the key concepts underpinning the operation of enhanced small cells is the separation of the control plane and the user plane. The control plane provides the connectivity and mobility while the user plane provides the data transport. In such a scenario, the user equipment (UE) will maintain connection with two different base stations, a macro and a small cell, simultaneously. The macro cell will maintain connectivity and mobility (control plane) using lower frequency bands, while the small cell provides high throughput data transport using higher frequency bands [1]. This is illustrated in Figure 1. An alternative version is the splitting of uplink and downlink across different classes of base stations.

The motivation behind this is that in the current 3GPP standard (Rel. 8-10), cell specific reference signals are always transmitted regardless of whether there are data to transmit or not, and transmitters cannot be switched off even when there is no data to transmit. However, with the definition of a new carrier type [3], where cell specific control signals, such as reference and synchronization signals, are removed, this is no longer the case. The macro cells will now provide the reference signals and information blocks, while the small cells, using the new carrier, can deliver data at higher spectrum efficiency, throughput, and energy savings. Additionally, they can now be switched off when there is no data to transmit. This can also provide additional benefits such as lower interference [4]. Such a scheme is expected to improve cell edge user throughput by up to 70 percent and reduce macro node energy consumption by 20 percent at low loads [4].

### Multiple Radio Access Technologies

Although the 3GPP define heterogeneous networks as the concurrent operation of different classes of base stations, we believe that heterogeneous networks in 5G will be a mixture of different radio access technologies as well. This will include future Wireless Local Area Network (WLAN) technologies which can offer seamless handovers to and from the cellular infrastructure, and device to device communications. This will lighten the burden on cellular networks and shift load away from the treasured licensed bands. At the same time, it can also concurrently provide higher throughput to users. This can already be implemented in part using the 3GPP Access Network Discovery and Selection Function (ANDSF) [5]. However, in situations where there is a high concentration of user terminals, offloading of data to WLANs may result in poor throughput, as WLANs are not well equipped to handle a large number of users. This problem is recognized by the IEEE 802.11 Working Group, which has initiated a study group on High Efficiency WLANs (HEW) to tackle situations where there is a high density of access points and/or a high density of user terminals.

### Device to Device Communications

Another approach to solving the highly dense network problem will be through Device to Device (D2D) communications, where each terminal is able to communicate directly with other terminals in order to either share their radio access connection, or to exchange information. Coupled with power control, D2D communications can reduce interference, especially in non-licensed frequency bands.

In 4G cellular communications, there are no provisions made for devices to communicate directly with nearby devices. All communications will have to be routed through the base station, and the gateway. This is extremely inefficient, especially when the devices are close by. In scenarios such as machine to machine (M2M) communications, where the number of devices involved can potentially be very large, it would be more sensible if devices can communicate directly with each other when necessary.

In unlicensed spectrum, devices can already communicate with each other outside of the cellular standard using technologies such as Bluetooth or Wireless LAN in *ad hoc* mode. However, these connections are susceptible to interference. On the other hand, using licensed spectrum will guarantee a certain level of quality of service if the connection is managed properly. These D2D communications will almost certainly require the base station to facilitate the connections to avoid intra-cell interference. The process to standardize this approach is already set in motion in 3GPP [6].

### Challenges of Heterogeneous Networks

**Inter-cell Interference** One of the biggest problems for HetNets is inter-cell interference. This is especially problematic with unplanned deployment of small cells, where the operators have little or no control of the location of the small cell. Additionally, the concurrent operation of small cells and traditional macro cells will produce irregular shaped cell sizes, and hence inter-tier interference, which will require advanced power control and resource allocation to avoid inter-cell interference.

**Distributed Interference Coordination** In deployment of access points where there are little or no coordination, such as between WLANs, distributed interference avoidance will be required. This will be increasingly crucial as more devices access unlicensed spectrum to complement their throughput.

**Efficient Medium Access Control** This is particularly relevant for dense deployment of access points and user terminals where the medium access is distributed, such as that of WLANs. In such situations, the user throughput is low, latency is high, and hotspots will not be able to complement cellular technology to provide a high throughput. Existing medium access control will need to be redesigned for such an environment to optimize the channel usage.

**Device Discovery and Link Setup** In non-network assisted device discovery in D2D communications, there could be issues when there is a large number of devices around. Additionally, setting up and maintain links with more than one party can prove to be difficult, especially when operating in the same frequency.

## 2. Software Defined Cellular Networks

In parallel with the development of software defined radio (or cognitive radio) in wireless communications, Software Defined Networking (SDN) has gathered momentum in the networking industry in the past few years. The concept of SDN originates from Stanford University's OpenFlow system [10], which enables abstraction of low level networking functionality into virtual services. In this way, the network control plane can be decoupled from the network data plane, which significantly simplifies network management and facilitates the easy introduction of new services or configuration changes into the network.

Currently in both academia and industry, a clear definition of SDN is still lacking. Nevertheless, according to a standardization body of SDN, Open Networking Foundation (ONF), the SDN architecture (as shown in Figure 2) has the following features [11]:

- Directly programmable: the network control plane is logically centralized and decoupled from the data plane. Network intelligence resides in software-based SDN controllers that maintain a global view of the network.
- Open: SDN simplifies network design and operation via open standards-based and vendor-neutral APIs (northbound and southbound).
- Agile: Network operators can dynamically configure, manage, and optimize network resources and adjust traffic flows to meet changing needs quickly via dynamic and automated SDN programs.

Recently, there are also growing interests in both academia and industry to apply SDN to mobile networks. The main motivation behind this is that SDN may help cellular operators simplify their network management and enable new services to support the exponential traffic growth envisaged for 5G networks. The authors of [12] argue that with open APIs and virtualization, SDN can separate the network service from the underlying physical infrastructure, thereby moving towards a more open wireless ecosystem and facilitating fast innovation. Similar to the programmable switches in wired SDN networks, programmable base stations and packet gateways are envisioned in cellular SDN architectures with extensions such as network virtualization on subscriber attributes and flexible adaptation of air interfaces [13]. Therefore, we believe that wireless or cellular SDN could be a possibility in future wireless networks.

Future 5G applications may have diverse characteristics and quality of service (QoS) requirements. For instance, M2M traffic has very different latency, throughput, and priority features compared to Human to Human (H2H) traffic. The same can be said for real-time video traffic and usual web browsing data traffic. The flexibility offered by SDN can enable fine-grained resource control (e.g. based on subscriber attributes) to enhance user Quality of Experience (QoE) while in the meantime maximizing network utilization [13].

At the moment, the wireless industry has yet to reach a consensus on a unified view of future 5G network architecture. Some favor a more distributed network architecture with self-organizing capability, while others have advocated more centralized cloud-based access networks (e.g. China Mobile's C-RAN). The development of cellular SDN is somewhat orthogonal to this ongoing evolution as it provides an open, flexible, and programmable middleware solution that can be used in different network architectures. Two important issues are scalability (to support a large number of small cells and a huge number of devices) and robustness (to provide a reliable abstraction without negatively impacting the flexibility).

### Challenges of Software Defined Networking

Wireless SDN is still at its infancy. There are a number of outstanding issues to be resolved before it can realize its full potential [14]. Firstly, further development is needed to apply SDN concepts at network infrastructure level, e.g. introducing SDN into carrier networks. For example, there is currently no consensus yet on how the programmable switch can be achieved in the best way (in terms of performance and flexibility trade-off). Secondly, global standardization is still on-going and a unified cellular programmable interface for implementing SDN infrastructures has yet to emerge. In particular, the standard development of a reliable Network Operation System (NetworkOS) that provides unified access to computing, storage and network resources is crucial for SDN implementation in a multi-vendor environment [14]. Finally, security in SDN is an open problem.

## 3. Massive MIMO and 3D MIMO

Another technology which is being considered is the use of a large array of antenna elements, several orders more than the number in use today, to provide diversity and compensate for path loss [7]. Otherwise known as Massive Multiple-Input/Multiple-Output (MIMO), it also allows for high resolution beamforming and is especially useful at higher frequencies where antenna elements can be miniaturized.

Massive MIMO can purportedly increase the capacity by several orders and simultaneously improve the radiated energy-efficiency [8]. In addition, it provides large number of degrees of freedom, which can be exploited using beamforming if the channel state information is available. Another advantage of Massive MIMO is its energy efficiency, and each antenna element is expected to use extremely low power [8].

However, there are several research challenges which need to be solved before Massive MIMO can be incorporated into future wireless systems. Beamforming will require a large amount of channel state information, and this will be problematic especially for the downlink. Consequently, Massive MIMO may be impractical for FDD systems, but can be used in TDD systems due to the channel reciprocity. Alternatively, limited feedback can be used. Additionally, Massive MIMO suffers from pilot contamination from other cells if the transmit power is high, and will suffer from thermal noise otherwise [8]. Last but not least, there is a lack of channel models for Massive MIMO systems, without which, researchers will not be able to accurately verify algorithms and techniques.

Another interesting technique currently considered is 3D MIMO, which allows for 3D beamforming. This is sometimes considered as a special type of large scale MIMO which is only concerned with using the antenna elements for beamforming. While normal beamforming methods form beams in two dimensions, 3D MIMO allows beam control in both horizontal and vertical directions. This additional control allows for further sectorization within a cell. An example of sectorization created by 3D MIMO is illustrated in Figure 3. As with Massive MIMO, 3D MIMO requires new channel models. Currently, 3GPP has started a work item on modelling 3D channels [9]. 3D MIMO will also require additional modifications to the feedback mechanism.

### Challenges of Massive/3D MIMO

**Channel Estimation/Feedback** Currently, only time division duplexing scenarios are considered for massive MIMO due to the prohibitive cost of channel estimation and feedback. Even for time division duplexing to work, channel calibration for Massive MIMO can prove to be a feat. New methods of channel estimation and feedback schemes will need to be proposed for massive MIMO to achieve mainstream status.

**Fast Processing Algorithms** To deal with the massive amount of data from the RF chains, extremely fast algorithms to process these data will be required.

**Pilot Contamination** Massive MIMO suffers from pilot contamination from other cells. Work around for this will be required for Massive MIMO to deliver its promised performance.

## 4. Machine to Machine Communications

As the enabling technologies described above continue to develop apace, fuelling the growth of service coverage and capacity, new use cases and applications are being identified, their emergence demanding yet more of our global networks. Many of these new business areas involve autonomous communication between devices, whether these devices are components in a smart energy network, intelligent home appliances or vehicles and infrastructure in an integrated transportation system. Indeed, we are already seeing examples of these "machine to machine" (M2M) devices: consider the latest generation of in-car satellite navigation ("sat-nav") devices with their integral cellular modems, downloading traffic information updates invisibly in the background. These new applications have the potential to cause a step-change in the size of the telecommunications market.

There are several challenges specific to M2M communications, not least of which are the autonomous operation and often restrictive power, size and complexity requirements. The typical M2M traffic is also quite distinctive: having spent the past decades optimizing our networks, from the highest level servers to the lowest level PHY channel codes, to support the characteristic traffic flows linked to speech, browsing and messaging, we are now faced with a different breed of traffic: short, periodic (or aperiodic) telemetry bursts and machine-generated updates. The distribution and nature of these M2M traffic flows do not sit readily within the current network architectures, so extensions and modifications are required.

The technology extensions developed and deployed to support these M2M applications have been appearing in a relatively *ad hoc* and piecemeal fashion, in specific standards bodies and/or organizations with particular technical or regional remits. Clearly, this vertical approach, while getting solutions out to market quickly, is not ideal in the long-term. There are areas of commonality in M2M solutions where consistent, standardized and open horizontal approaches will help develop the economies of scale and interoperability that will lead to a truly global M2M market.

It was to this end that the "oneM2M" Partnership Project (PP) was formed during 2012, to develop global, access-technology agnostic Service Layer specifications for M2M, in the same mold as 3GPP. This international body was formed by Standards Development Organisations (SDOs) from across the world: ETSI from Europe, ATIS and TIA from the USA, TTC and ARIB from Japan, CCSA from China and TTA from South Korea. Each of these SDOs already had interests in (and, in many cases, solutions for) different aspects of, and variations on, M2M systems: for example, the ETSI Technical Committee M2M had already produced an entire tranche ("Release 1") of technical requirements and specifications. These existing standards range from architectural descriptions to interface definitions, such as service layer interactions with common cellular access systems such as those developed by 3GPP and 3GPP2 (who are already developing and releasing extensions to their recommendations to support M2M traffic).

After this first year, the oneM2M participants (delegates from 200+ participating member companies and interested parties) have compared, merged, down-selected and harmonized "best of breed" contributions and proposals from around the world. The biggest challenges have often been finding common ground and vocabulary between different proposals.

The architecture is based around Common Service Functions (from Device Management to Session Management) residing within Common Service Entities (CSE), with interfaces between the CSE and the applications above and the underlying network services below clearly defined. The current feeling is that both service- and resource-orientated architectures on the key CSE interfaces should be supported.

The system is designed adopting the REST philosophy (a "RESTful" system). That is, the system is stateless, with uniquely addressable entities. Furthermore, the system must have well-defined interfaces between client and server, and between layers, to allow independent development and evolution of components. This RESTful approach is a key enabling technology for not only the "Internet of Things" but also the longer-term "web of things", which we believe will be an application within 5G systems.

Communication flows based around request/response interactions are also defined, and protocols for the different interfaces are being identified and scoped.

The key, top-level documents have already been finalized and agreed, and are now under change control processes. Work continues within the different Working Groups to finalize and agree the remainder. The group is anticipated to deliver a first release in the middle of 2014.

### Challenges of M2M
Specific challenges vary according to the exact M2M application, but there are two themes that recur in many, if not all, M2M applications:

**Massive access**: Compared to conventional human to human traffic in cellular networks, a huge number of M2M devices in a cell can pose serious system challenges in terms of radio access network (RAN) congestion and overload. Currently a number of proposals have been proposed in 3GPP to address the RAN overload issue, e.g. back-off adjustment, access class barring, and M2M prioritization. However, each of these methods has its strengths and weaknesses and none of them is widely acknowledged as the best solution.

**Security and privacy:** Security has been widely discussed in various standardization bodies. For instance, in ETSI M2M [15], M2M security focuses on several attributes of a user and their communications, including authenticity, authority, integrity, and confidentiality. To enable wide deployment of M2M services and especially enhance consumer acceptance, M2M privacy is of paramount importance. Different M2M applications and sectors (e.g. e-health and smart metering) may have different privacy requirements which have to be taken into account right from the beginning of system design.

## 5. Other Technologies
Apart from the above technologies and applications, the following technologies can also potentially impact 5G.

### Millimetre Wave
An obvious way of increasing the throughput will be through bandwidth expansion. However, the available bandwidth below 6 GHz is limited, and re-farming analogue TV spectrum will not sufficiently meet the burgeoning demand. Already, there are efforts to look beyond 6 GHz and also at the millimetre wave frequencies to evaluate their feasibility for use in future networks. However, the characteristics of higher frequencies are not well studied, and measurement campaigns and channel modelling for different scenarios and environments will be required before transmission technologies can be designed for them. We believe that millimetre wave frequencies holds the most promise, and there are already on-going efforts to make this a possibility. In [15], millimetre wave frequencies of 28 GHz and 38 GHz are extensively studied to understand their propagation characteristics in different environments, paving the way for their use in future wireless systems.

### Shared Spectrum
Although cognitive radio was often touted as a solution to the problem of frequency spectrum shortage, it is seldom adopted as there are always concerns about the impact on the primary user or license holder of the spectrum. An alternative solution proposed which can potentially solve this dilemma is Authorized Spectrum Access (ASA) also known as Licensed Spectrum Access (LSA) [17]. The concept of LSA is to allow authorized users to access licensed spectrum based on certain conditions set by the licensee of the spectrum. This would allow under-utilized spectrum to be more effectively used and also solve the problem of quality of service for the primary user.

### Big data
Like in many other market sectors and industries, big data will also bring about lots of challenges and opportunities in 5G wireless. First of all, cellular networks have to provide efficient infrastructure support for this data deluge. For example, the future M2M or Internet of Things (IoT) applications will generate a vast amount of data. As discussed previously, this proves to be a major technical challenge for RANs. Secondly, new network architectures may emerge from the necessity of running big data applications. There is close synergy between cloud computing, software defined networking, and Network Function Virtualization (NFV). A convergence of these technologies can be envisaged to form highly robust and reliable 5G platforms for big data. Thirdly, making informed decisions and extracting intelligence from big data is an extremely important and yet non-trivial task. For example, cellular operators can make use of various customer network access data to reduce churn rate and seek new revenue opportunities. The smart grid, as another example, can be seen as a huge sensor network, with immense amounts of grid sensor data from various sensors, meters, appliances and electrical vehicles. Data mining and machine learning techniques are essential for efficient and optimized operation of the grid.

### Indoor Positioning
While indoor positioning itself does not improve throughput or coverage, it has large implications on various applications and the quality of communications. Accurate positioning of user terminals can provide the network with additional information that can help in resource allocation and quality of service improvement. It can also enable a plethora of applications, including position based handover, resource allocation, and location based services.

Currently, 3GPP LTE has several positioning methods, including Cell ID (CID) and Enhanced Cell ID (ECID), as well as Assisted Global Navigational Satellite Systems (A-GNSS). It is also able to position using the Observed Time Difference of Arrival (OTDOA) method. All these are enabled through the Enhanced Serving Mobile Location Centre (E-SMLC) using the LTE Positioning Protocol (LPP) [18]. Accuracy improvements to the currently available methods will certainly open opportunities for more location based applications.

## Conclusions

In this paper, we have provided an overview of some emerging technologies which may make up future 5G wireless networks. We have also described some research problems which these technologies present.

While there is currently no clear consensus among academics and industrials on what will define 5G wireless networks, we believe that future 5G wireless networks will be a combination of different enabling technologies, and the biggest challenge will be to make them all work together.

# Biographies

WOON HAU CHIN [SM] (w.h.chin@ieee.org) is a principal research engineer and team leader at Toshiba Research Europe Limited in Bristol, United Kingdom. He was formerly with the Institute for Infocomm Research (I2R) in Singapore. He holds a B.Eng (1$^{st}$ class honours) and a M.Eng in electrical engineering from the National University of Singapore, and a Ph.D. in electrical and electronic engineering from Imperial College London. He has contributed to multiple standards, including IEEE 802.11n and 3GPP LTE. His research interests are statistical signal processing and smart grid communications.

ZHONG FAN (zhong.fan@toshiba-trel.com) is a chief research fellow with Toshiba Research Europe in Bristol, United Kingdom. Prior to joining Toshiba, he worked as a research fellow at Cambridge University, a lecturer at Birmingham University, and a researcher at Marconi Labs Cambridge. He was awarded a BT Short-Term Fellowship to work at BT Labs. He received his B.S. and M.S. degrees in electronic engineering from Tsinghua University, China, and his Ph.D. degree in telecommunication networks from Durham University, United Kingdom. His research interests are 5G wireless networks, big data, M2M, and smart grid communications.

RUSSELL J. HAINES [SM] (russell.haines@toshiba-trel.com) is the Chief Laboratory Coordinator at Toshiba Research Europe's Telecommunications Research Laboratory in Bristol, United Kingdom. He holds a first-class B.Eng.(Hons) degree and a Ph.D. from the University of Bristol. Before joining Toshiba he worked on GSM (2G) mobile phone development for NEC and on an emergency services' communications switch at GEC Marconi. His research interests include reliability and connectivity of networks; practical applications of formal mathematical methods; interference avoidance and management; and alternative mode monitoring for software defined radio and cognitive radio. He has contributed to standards, papers, and products in the smart energy management, machine-to-machine, mobile/cellular (including small/femto cells), WLAN (esp. IEEE 802.11n and HEW), and WPAN (e.g., Bluetooth) fields. He holds over 30 patents. In addition to being a Senior Member of the IEEE, he is a Visiting Fellow at the University of Bristol, a Fellow of the Institution of Engineering and Technology, a Chartered Engineer, a Eur Ing (European Engineer), and a Senior Member of ACM.

# Tables

TABLE 1: DIFFERENT ENABLING TECHNOLOGIES

|  | Energy Efficiency | Capacity Enhancement | Coverage |
|---|---|---|---|
| Massive/3D MIMO | ✓ | ✓ |  |
| Dense HetNets | ✓ | ✓ | ✓ |
| Multi-RAT Technologies |  | ✓ |  |
| D2D | ✓ |  | ✓ |

TABLE 2: TAXONOMY OF ENABLING TECHNOLOGIES

| *Radio access | New network architecture | New applications |
|---|---|---|
| Massive/3D MIMO | HetNets | M2M |
| Multi-RAT | SDN | Localization and positioning |
| D2D |  |  |
| Millimetre wave |  |  |

# Figures

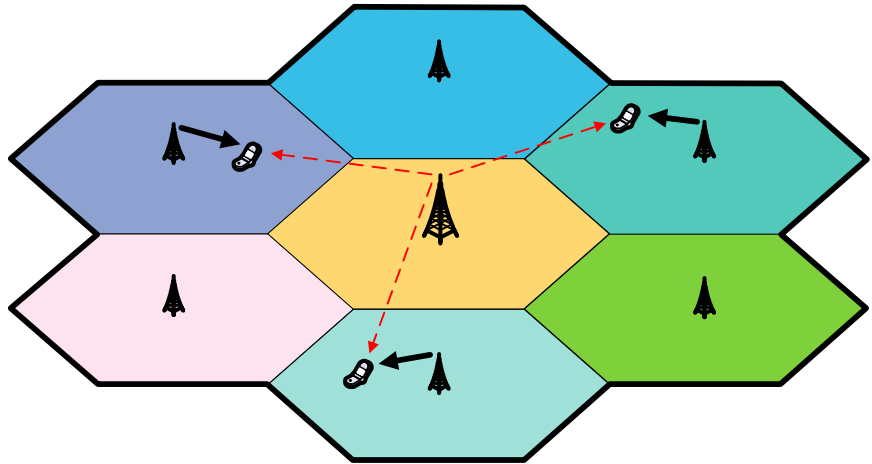

FIGURE 1: SEPARATION OF CONTROL AND USER PLANES

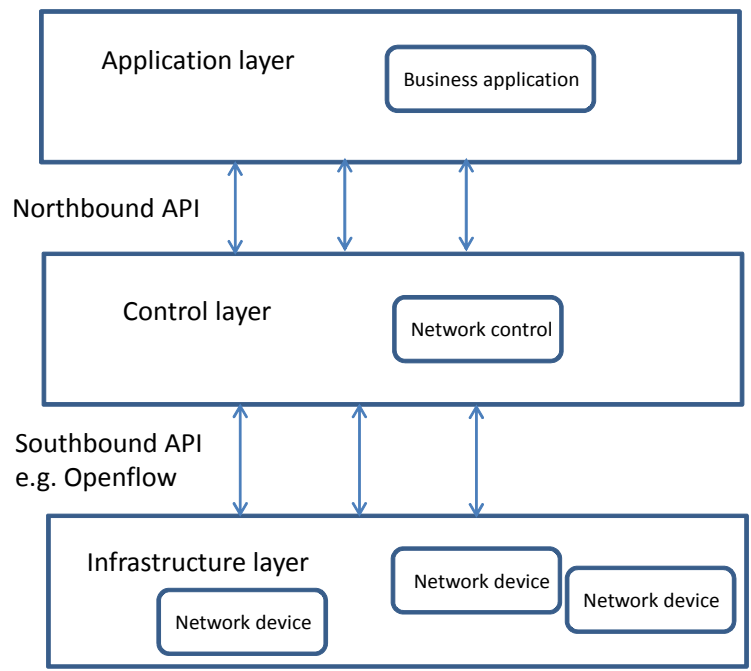

FIGURE 2: THE SDN ARCHITECTURE

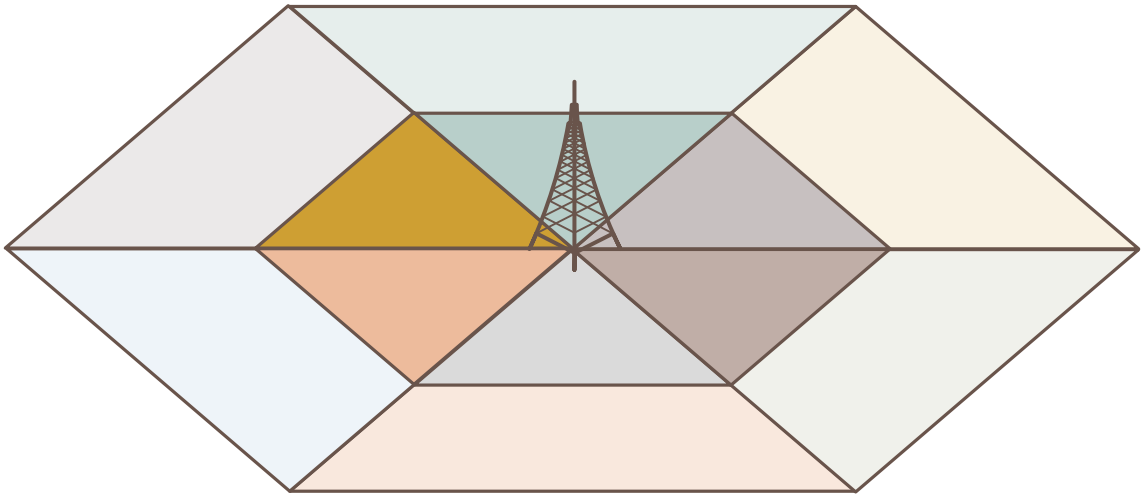

Figure 3: Possible Sectorization for 3D-MIMO